\documentclass[nature,article,onecolumn,preprintnumbers,amsmath,amssymb,superscriptaddress]{revtex4}

\date{\today}
\usepackage{epsfig}
\usepackage{subfigure}
\usepackage{amsmath,amssymb}
\usepackage{graphicx}
\usepackage{dcolumn}
\usepackage{bm}
\usepackage{bbm}
\usepackage[colorlinks,linkcolor=blue,hyperindex,CJKbookmarks]{hyperref}
\usepackage{float}
\usepackage{hyperref}
\usepackage{mathtools}
\usepackage{physics}
\usepackage{comment}
\usepackage{color}
\allowdisplaybreaks

\setcitestyle{numbers,square}

\usepackage[normalem]{ulem}
\usepackage{setspace}
\begin{document}

\title{Observing photo-induced chiral edge states of graphene nanoribbons in pump-probe spectroscopies}
\author{Yuan Chen}
 \affiliation{Department of Applied Physics, Stanford University, Stanford, California 94305, USA}
 \affiliation{Stanford Institute for Materials and Energy Sciences, SLAC National Accelerator Laboratory and Stanford University, Menlo Park, California 94025, USA}
\author{Yao Wang}
 \email[Email: \href{mailto:yaowang@clemson.edu}{yaowang@clemson.edu}
]{}
\affiliation{Department of Physics and Astronomy, Clemson University, Clemson, South Carolina 29634, USA
 }
\author{Martin Claassen}
\affiliation{Center for Computational Quantum Physics, Simons Foundation Flatiron Institute, New York, NY 10010, USA}
\author{Brian Moritz}
\affiliation{Stanford Institute for Materials and Energy Sciences, SLAC National Accelerator Laboratory and Stanford University, Menlo Park, California 94025, USA}
\affiliation{Department of Physics and Astrophysics, University of North Dakota, Grand Forks, North Dakota 58202, USA}
\date{\today}
\author{Thomas P. Devereaux}
 \email[Email: \href{mailto:tpd@stanford.edu}{tpd@stanford.edu}
]{}
\affiliation{Stanford Institute for Materials and Energy Sciences, SLAC National Accelerator Laboratory and Stanford University, Menlo Park, California 94025, USA}
\affiliation{Department of Materials Science and Engineering, Stanford University, California 94305, USA}
\affiliation{Geballe Laboratory for Advanced Materials, Stanford University, Stanford, California 94305, USA}
\begin{abstract}
Photo-induced edge states in low dimensional materials have attracted considerable attention due to the tunability of topological properties and dispersion. Specifically, graphene nanoribbons have been predicted to host chiral edge modes upon irradiation with circularly polarized light. Here, we present numerical calculations of time-resolved angle resolved photoemission spectroscopy (trARPES) and time-resolved resonant inelastic x-ray scattering (trRIXS) of a graphene nanoribbon. We characterize pump-probe spectroscopic signatures of photo-induced edge states, illustrate the origin of distinct spectral features that arise from Floquet topological edge modes, and investigate the roles of incoming photon energies and finite core-hole lifetime in RIXS. With momentum, energy, and time resolution, pump-probe spectroscopies can play an important role in understanding the behavior of photo-induced topological states of matter.
\end{abstract}

\maketitle
\thispagestyle{plain}
\section{Introduction}
The quest for controlled manipulation of quantum states of matter with light promises to reveal and ultimately functionalize properties of materials far from equilibrium, while posing profound theoretical and experimental challenges in probing and understanding the underlying microscopic dynamics. Fundamentally, irradiating materials with light permits selectively changing electronic distributions and altering the energetics of states that couple to the light, inducing varieties of nonequilibrium phases\cite{Cavalleri2016,Light-inducedSuperconduct,lee2012phase,kim2012ultrafast,gerber2015direct,wang2016using,ZX2012,ZX2008,Sentef2015}. 
Driven by the search for Majorana fermions and applications in quantum computing\cite{RevModPhys.82.3045}, transient photo-induced topological band insulators and superconductors have recently garnered much attention \cite{oka2009photovoltaic,kitagawa2010topological,kitagawa2011transport,FTI2011Galitski,FTI2013photonic,grushin2014floquet,dehghani2014dissipative}. These rely solely on transient modifications of the single-particle band structure. Different from static topological insulators, such Floquet topological insulators (FTIs) can be tuned via amplitude, frequency, and polarization of the pump light, making them versatile and easier to control. 

Graphene irradiated with circularly-polarized light constitutes a paradigmatic example of an FTI. Pristine graphene enjoys time-reversal and inversion symmetry, and has massless Dirac fermions at $\rm K$ and $\rm K'$. Haldane predicted that breaking time-reversal symmetry opens a gap at the Dirac points, triggering a transition to a Chern insulator\cite{haldane1988model}. A natural nonequilibrium realization follows from pumping graphene with circularly polarized pump light, which induces chiral edge modes at the sample boundary that span the Floquet bandgap of the photon-dressed electronic system \cite{gu2011floquet,perez2014floquet,usaj2014irradiated,dehghani2015out}, and it has been shown that the topology of an irradiated graphene nanoribbon can be tuned either via the pump frequency and/or amplitude\cite{mikami2016brillouin,wang2016edge}. 
Previous characterizations of Floquet topological states in graphene nanoribbons have mainly focused on transport properties\cite{oka2009photovoltaic,gu2011floquet,dehghani2015out,zhai2014photoinduced,puviani2017dynamics,torres2014multiterminal,farrell2016edge}, especially Floquet generalizations of the quantum anomalous Hall effect. 
For example, McIver \textit{et al.}\cite{mciver2020light} observed the light-induced anomalous Hall effect in graphene under circularly polarized light. They observed a plateau of Hall conductance when the Fermi energy lies within the light-induced gap, but theoretical calculations\cite{sato2019microscopic} suggest a population imbalance of carriers -- in addition to a change of topology -- as the root causes of this effect. Theoretically, Foa Torres \textit{et al.}\cite{torres2014multiterminal} predicted that the DC quantum Hall conductance is not directly linked to topological invariants of the full Floquet bands - not all Floquet edge modes contribute equally to the Hall conductance. These all necessitate a direct and conclusive technique to detect Floquet topological states.

As an advanced spectroscopic technique that maps the single-particle dispersion and occupation with momentum, energy, and time resolution\cite{FreericksPRL2009, wang2018theoretical}, trARPES has directly observed photo-induced Floquet-Bloch states on the surfaces of topological insulators\cite{Gedik2013Science,NuhGedik2016NP}, and has been predicted to allow detection of photo-induced exotic phases in graphene\cite{Sentef2015}. Though powerful in characterizing instantaneous band structures, trARPES requires premium sample quality, high vacuum, and a clean electromagnetic environment, limiting its applicability to materials in extreme conditions. In contrast, trRIXS is a photon-in-photon-out spectroscopic technique without much restriction on sample and environmental conditions. Recent trRIXS experiments have revealed the anisotropy in the dynamics of magnetic excitations in $\rm Sr_2IrO_4$\cite{Dean2016trRIXS,cao2019ultrafast} and transverse charge fluctuations in $\rm La_{2-\mathnormal{x}}Ba_\mathnormal{x}CuO_4$\cite{mitrano2019ultrafast}. Access to energy-momentum-resolved collective excitations makes trRIXS an indispensable tool for characterizing the properties of nonequilibrium or driven quantum materials\cite{TPD2016}. 

In this paper, we demonstrate that trRIXS can provide direct state-selective evidence of the induction of transient edge modes. Remarkably, we show that the state selectivity of trRIXS entails that photo-induced chiral edge modes feature prominently in the multi-particle channel when the intermediate state has a relatively long lifetime. Combined with tuning of the incoming photon energy in trRIXS, this permits selective experimental probes of two classes of Floquet edge modes\cite{rudner2013anomalous}, which either bridge the photo-induced gaps at the Dirac point or induced hybridization gaps at energies $\pm \Omega/2$ away from the Dirac point, as a function of pump photon frequency $\Omega$. Our results demonstrate a way for a direct and conclusive experimental characterization of Floquet edge modes in FTIs, in graphene and beyond.

\section{Results and Discussion}
\subsection{Pump-Probe Spectroscopies For High Pump Frequency}\label{sec:pumpProbeRes}

We present trARPES and trRIXS calculations for a 60-atom wide zigzag graphene nanoribbon under a circularly polarized pulsed laser pump. The nanoribbon is depicted in Fig.~\ref{fig:1}, while the details of the model and the pump mechanism are explained in the Methods section. We first discuss the high-frequency pump $\Omega=6.2t_h$ in this section, which is off-resonant since it is larger than the equilibrium bandwidth ($6t_h$). 


Figure~\ref{fig:2}(a)-(e) show the trARPES spectra in the presence of the high-frequency pump. Before the pump pulse, the trARPES spectrum shows occupied electronic states at zero temperature, including the lower half of the Dirac cone (corresponding to $k=\pm 2\pi/3$) and the $\omega=0$ edge states at large momenta. As the pump pulse turns on, it transiently manipulates the single-particle states. The bandwidth of bulk states is renormalized due to Floquet photon dressing. At the same time, the edge states in the nanoribbon system\cite{wakabayashi2010electronic} become dispersive as a consequence of gap opening at the Dirac points due to time-reversal symmetry breaking. The nonequilibrium band structure can be quantitatively obtained from the Floquet theory, by approximating the pump with a periodic oscillation with the same instantaneous pump amplitude $\bar{A}(t)$ [see the Supplementary Note 2]. With the off-resonant pump, after the pump pulse, the trARPES spectrum returns to its equilibrium value since the pump cannot inject any resonant excitations. 

The trRIXS spectrum requires the incoming photon energy $\omega_{\rm i}$ tuned close to the atomic $K$-edge energy $E_0$. We consider the $\omega_{\rm i}=E_0$ condition and core-hole lifetime $\sigma_{\rm ch}=\textrm{5.73fs}$\cite{sette1990lifetime}, and examine a full-range of transferred momentum in this section, postponing discussions of resonances and core-hole lifetimes to Sec.~\ref{discussion}. Figure~\ref{fig:2}(f)-(j) gives an overview of the trRIXS spectra at five different times corresponding to the trARPES snapshots. Before the pump pulse enters ($t=-4\sigma_{\rm pu}$), the equilibrium RIXS spectrum roughly depicts the charge excitations. The particle-hole continuum starts at zero momentum and energy, reflecting the scatterings of electrons near the Dirac points. This continuum becomes gapless at $q=4\pi/3$ due to scattering between the two Dirac points. In addition to this, distinct features of zigzag graphene nanoribbons are the flat excitations near $q=0$ and $q=4\pi/3$ (magnified in the red and pink zoom-in view respectively), which correspond to trivial bound states at the edge. 

In the presence of the pump field, a gap opens near the Dirac points, as previously shown by trARPES. In trRIXS, the gap is reflected in the hardening of the bottom of the spectral continuum at $q=0$ and $q= 4\pi/3$ [see Figs.\ref{fig:2}(g)-(i)]. Simultaneously, the flat edge excitations become dispersive, signifying the appearance of chirally-propagating edge modes. Furthermore, trRIXS also exhibits a softening at the top of the spectrum, altogether leading to a squeezing of the compact support. This is a consequence of the Floquet band renormalization, which reduces the single-particle bandwidth by around $\mathcal{J}_0[\bar{A}(t)]$\cite{YaoNoneq,mypaper}. 

After the pump pulse, the trRIXS spectrum recovers following the reversal of time, as shown in Fig.~\ref{fig:2}(j). Note that the recovery is not guaranteed in general because a generic pump pulse can inject energy and particle-hole excitations to the system, resulting in changes of electron occupations after the pump. However, here due to the off-resonance of the pump field and the lack of interaction, states within the same sideband have a negligible transition rate. Thus the electrons cannot reach the initially empty states, and no remnant excitations exist in this system after the pump finishes.  

We want to acknowledge that experimentally, it is difficult to generate and manipulate a pump pulse with the photon energy as high as 16.7eV. The case here just exemplifies the observation of the gap opening and chiral edge states at the Dirac points. In Supplementary Note 5 and Supplementary Fig.~3, we study the trARPES and trRIXS for a frequency pump of 2eV. Similar results can be found there.

\subsection{Analysis and Discussion For TrRIXS Spectra}\label{discussion}

To better understand the origin of the new features emergent in the pumped graphene nanoribbon, we apply an extra spatial filter in the numerical calculation through a Gaussian envelope whose center is denoted by $y_0$ [see Methods for details]. Physically, the filter highlights RIXS signals from the bulk or edge of the nanoribbon, selectively tuned as one changes the center $y_0$. While still experimentally inaccessible, many x-ray sources including Linac Coherent Light Source (LCLS) at SLAC National Accelerator Laboratory are actively developing the `nano-RIXS' technique such that focused local probing will be possible in the near future. Figure~\ref{fig:3} (a)-(c) show the filtered trRIXS signals from the upper edge, lower edge, and ribbon center respectively. We recognize that different parts of the nanoribbon are responsible for different features in the spectra. The upper edge contributes to a linear low energy mode with a positive slope near $q=0$, while the lower edge contributes to a similar mode but with a negative slope near $q=2\pi$. The central part of the ribbon is gapped, and there are no excitations below $\omega\approx 0.45t_h$. These results verify that the linear low-energy modes near $q=0$ and $q=2\pi$ are indeed edge modes, while the bulk nanoribbon is gapped. A more quantitative analysis comparing to Floquet single-particle spectrum can be found in the Supplementary Note 4 and Supplementary Fig.~2.


To reveal the origin of the edge features, we take advantage of state selectivity in RIXS and vary the incoming photon energy $\omega_{\rm i}$ near the absorption edge $E_0$.
Figure~\ref{fig:4} shows the trRIXS spectra in a range of small energy loss $\Delta\omega$ for different $\omega_{\rm i}$'s. To guide the eye for the corresponding intermediate states, we also denote the positions of these $\omega_{\rm i}$'s in the Floquet spectrum in Fig.~\ref{fig:4}(e). For $\omega_{\rm i} =E_0-0.34t_h$, the photo-electron in the intermediate state does not have enough energy to occupy any available single-particle state, leading to a blank trRIXS spectrum. With the increase of $\omega_{\rm i}-E_0$ above the valence band, the photo-electron overlaps with the unoccupied density of states, giving finite trRIXS spectral weight. The $\omega_{\rm i}-E_0$ can be further divided into three different ranges: when $\omega_{\rm i}-E_0$ is within the gap of the bulk Floquet spectrum, the photo-electron can only occupy the edge states, resulting in an evident edge feature [see Fig.~\ref{fig:4}(b)]; when $\omega_{\rm i}-E_0$ reaches the conduction band, the high-energy particle-hole excitations across the bulk bandgap start to appear in the trRIXS spectrum [see Fig.~\ref{fig:4}(c)]; finally, with even larger $\omega_{\rm i}$, the photo-electron can no longer stay in an edge state, and the edge features gradually disappear in trRIXS [see Fig.~\ref{fig:4}(d)]. Particularly, since $\omega_{\rm i}-E_0$ goes farther beyond the gap, the bulk excitations are bounded by the energy difference between the intermediate state and the top of the valence band and therefore hardens beyond the given range of $\Delta\omega$. Videos of the trRIXS snapshots as $\omega_{\rm i}$ sweeps can be found in the Supplementary Videos.
Therefore, using the state selectivity provided by the incoming photon energy, we can further associate the edge features in trRIXS with specific single-particle states. In Sec.~\ref{low-energy}, we will show how this property helps decipher different kinds of edge states for low pump frequency.


Since the state selectivity of trRIXS stems from the resonant intermediate state, the edge features should depend on the core-hole lifetime. In RIXS, the interpretation of spectral intensity has been restricted to dynamical charge or spin structure factors, where the cross-section is simplified by assuming an ultrashort core-hole lifetime (UCL)\cite{van2005theory,van2005correlation,ament2007ultrashort, YaoNoneq}. This is reasonable for properties where the details of the intermediate state may be less relevant. However, as shown in Fig~\ref{fig:4}(f), the edge features are almost invisible in the dynamic structure factor $N(\mathbf{q},\omega,t)$ compared to the intense bulk features. Instead, if we increase the core-hole lifetime $\sigma_{\rm ch}$ to finite values ($\textrm{1.15}{\rm fs}$ and $\textrm{5.73fs}$), the edge features become more apparent, relative to the bulk [see Figs.~\ref{fig:4}(g) and (h)]. Since 1D edge states would naively be expected to have a small cross section compared to two-dimensional bulk states, the spectral weight associated with edge states is usually dwarfed by the bulk signal in both ARPES and $N(q,\omega)$. However, via appropriate choice of the incident photon energy, one can selectively highlight intermediate states, which are connected to final states dominated by the Floquet edge state. A longer core-hole lifetime increases the impact of this topological intermediate state to the entire cross-section and enhances the edge features in the trRIXS spectrum. Therefore, the capability of trRIXS in resolving topological edge states lies in its nonlinearity beyond the UCL approximation. Similar core-hole lifetime induced nonlinear effects have been discussed in equilibrium RIXS of correlated materials in terms of bimagnon excitations\cite{ChunjingPRX,tohyama2018spectral}.

\subsection{Detecting Different Edge Modes at Low Pump Frequency}\label{low-energy}
Previous studies of the Floquet single-particle spectrum suggested that topological edge modes of a graphene nanoribbon reside in two possible kinds of band gaps: either the Dirac point gaps at energy $n\Omega$ or the dynamical band gaps at energy $(n+1/2)\Omega$\cite{usaj2014irradiated}. While it is hard to directly characterize the latter ones in transport experiments\cite{torres2014multiterminal}, we expect that trRIXS can distinguish these edge modes with momentum resolution. Since these edge modes are absent for a non-resonant pump, we switch to a smaller pump frequency $\Omega=3t_h=8.1\rm eV$ in this section, which is smaller than the equilibrium bandwidth. Other pump and probe conditions remain the same as Sec.~\ref{sec:pumpProbeRes}. As shown in Fig.~\ref{fig:5}(a), the new pump results in edge states in the dynamical gap at around $\omega=1.5t_h$, different from the one arising at the two Dirac points. The electron occupation of these single-particle states also becomes more complicated than the case of the non-resonant pump [see the Supplementary Note 2 and Supplementary Fig.~1 for detailed discussions]. 

To distinguish the different edge states, we tune the incoming photon energy in the trRIXS and take advantage of the selectivity of the intermediate states [see Figs.~\ref{fig:5}(b)-(d)]. For $\omega_{\rm i}=E_0$, linear dispersions rise near $q=0$ and $q=2\pi$ similar to Fig.~\ref{fig:2}(h), corresponding to the edge state near the Dirac points in Fig.~\ref{fig:5}(a). With larger $\omega_{\rm i}$ in Fig.~\ref{fig:5}(c), the edge features are no longer visible, leaving the spectrum dominated by incoherent spectral continuum at low energy. This is because at $\omega_{\rm i} = E_0+t_h$, the excited electron in the intermediate state lies in the continuum of bulk states. For even larger $\omega_{\rm i}=E_0+1.5t_h$, the trRIXS displays linear edge features again. Different from Fig.~\ref{fig:5}(b), there are two extra edge features dispersing from $q=\pi$. These stem from the particle-hole excitations between the edge states in a small dynamical gap (about $0.2t_h$) selected by the incoming photon energy. More specifically, the scattering between left- and right-moving edge states leads to an offset of $\sim\pi$ in momentum and correspond to the features near $q=\pi$, while the scattering within the same edge state gives the linear features near $q=0$ and $2\pi$, similar to Fig.~\ref{fig:5}(b). 
In addition to the edge states, one can also observe a feature starting at $\Delta\omega\approx 0.2t_h, q=0$ which indicates a bulk gap of size $0.2t_h$. Note that real excitations exist under a resonant pump. In Fig.~\ref{fig:5}(a) and (c), $\omega_{\rm i}$ lies at the center of the gaps, and the spectral contributions from bulk particle-hole excitations are suppressed by the x-ray resonance. But for Fig.~\ref{fig:5}(b), those excitations can be reached by the intermediate state whose energy corresponds to bulk states in a continuum. Therefore, trRIXS can characterize the collective excitations associated with both the bulk and two different kinds of edge states through the control of incident photon energy. When the dynamical gap is smaller than the resolution of trARPES, trRIXS provides a way to detect these edge states. 

\subsection{Conclusion and Outlook}
In conclusion, we numerically predicted trARPES and trRIXS spectra for a zigzag graphene nanoribbon under circularly polarized irradiation. These spectroscopic probes directly map the electronic states in pumped graphene nanoribbons as well as the particle-hole excitations with time, momentum, and energy resolution. We further decomposed the spectral contributions from the upper edge, the lower edge, and the bulk by adding local filters. We also investigated the influence of incoming photon energy and core-hole lifetime to the trRIXS spectra and showed that they are crucial for selecting the excitations with corresponding intermediate state energy, which is particularly useful for separating different kinds of edge modes when the pump frequency is low. These results illustrate methods to detect and understand the behaviors of pumped graphene nanoribbon.

In previous trARPES studies of graphitic materials, people are mainly interested in the electron relaxation dynamics after the pump creates an electron-hole pair. The relaxation process includes complicate combinations of electron-phonon coupling as well as Auger heating and impact ionization caused by electron-electron scattering that are still not fully understood\cite{PhysRevLett.115.086803,PhysRevLett.121.2564013,jphyc,bilayer_arpes}. While phonons dominate around 250fs after the pump when hot Fermi-Dirac distributions of electrons and holes establish\cite{relax}, both electron-phonon coupling and electron-electron scattering are involved in the relaxation almost immediately after the excitation. In those trARPES work, low energy excitations near the Dirac points can be observed as a result of such relaxation. In our work, to highlight the spectral features from chiral edge states with the easiest possible model, we have neglected the phonons, electron-electron interaction, and matrix elements for transitions (see Methods for details of the model). For a graphene nanoribbon under a circularly polarized light, the photo-induced gap at the Dirac points will likely suppress impact ionization that produces low-energy particle-hole pairs near the Dirac point, because an extra energy has to be paid to form such excitations in a gapped system. When the x-ray is tuned resonant within the gap, the signals from possible bulk excitations induced in the relaxation process will be reduced as a result of selectivity from resonant intermediate states. Since our model lacks electron-phonon coupling and electron-electron scattering, their specific role in the trRIXS of the graphene nanoribbon will need to be investigated in future studies. While RIXS can observe electron-phonon coupling\cite{TPD2016,phonon_theory,yingying_2020} and excitations caused by electron correlation\cite{my___paper}, in the case of the graphene nanoribbon it is unclear whether they will have significant contributions to the spectra\cite{Oppen}.

In this paper we only numerically evaluated the spectra for a zigzag nanoribbon at a fixed width. While armchair nanoribbons do not hold edge states at equilibrium, they will still hold chiral edge states when pumped by circularly polarized light, because they share the same underlying bulk topology as the zigzag nanoribbons\cite{rudner2013anomalous,usaj2014irradiated}. However, the dispersion is not guaranteed to be the same. The width of the nanoribbons will not have a significant influence on the spectra, except when it is so narrow that the edge states on different edges mix up. The experimental scheme described in this paper applies to either an isolated graphene nanoribbon or an ensemble of aligned graphene nanoribbon (like the samples in \cite{acsnano}). In the ensemble case, the final signal will be an incoherent sum of signals from different ribbons.

We stress that the pump-probe methods studied in this paper are not limited to graphene nanoribbon; the same trARPES and trRIXS experiments will also apply to other topological materials (e.g. transition metal dichalcogenide (TMDC) under circularly polarized light\cite{claassen2016all}). Since carbon has a shallow x-ray edge energy (around 285eV\cite{PhysRevLett.40.1296}) which could not cover enough range of the Brillouin zone, the graphene nanoribbon is not the best material for tr-RIXS. TMDC, on the other hand, are better choices because of higher x-ray edges (e.g. S $K$-edge or Se $L$-edge). We believe this paper will guide future experiments to conclusively characterize photo-induced topological states of matter.

\appendix

\section*{Methods}\label{sec:model}
\subsection{Model of the Graphene Nanoribbon}

\begin{figure}
\includegraphics[width=8.5cm]{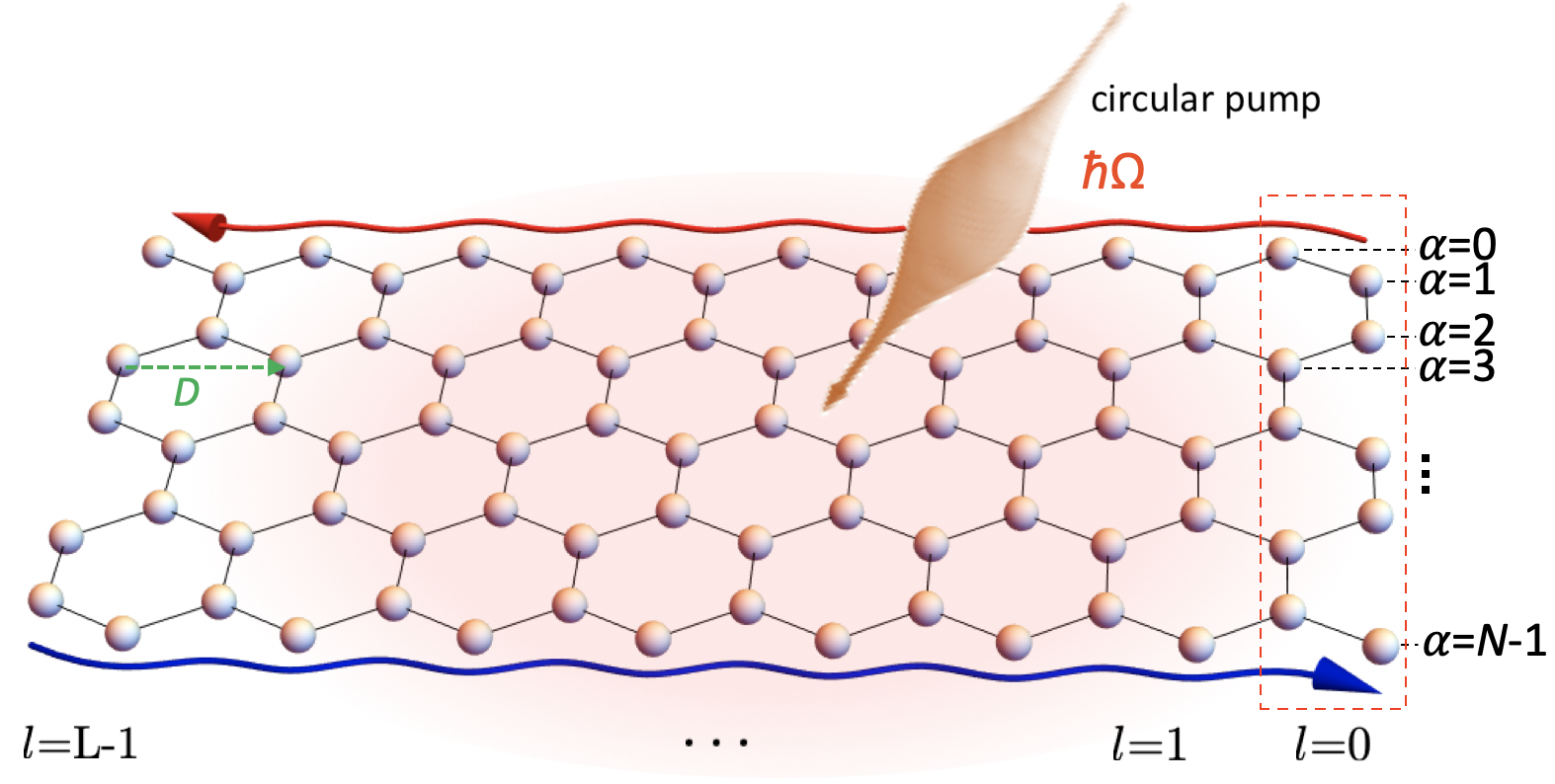}
\caption{\label{fig:1} \textbf{Schematics of an irradiated nanoribbon.} A ribbon with a width of $N$ atoms is irradiated via a circularly-polarized pulse. The unit cell is labeled by $l$ and the atomic position within a cell is denoted via $\alpha$. The ribbon unit cell width is $D = \sqrt{3}a_0$ with $a_0 $ the bond length.}
\end{figure}

\begin{figure}
\includegraphics[width=18cm]{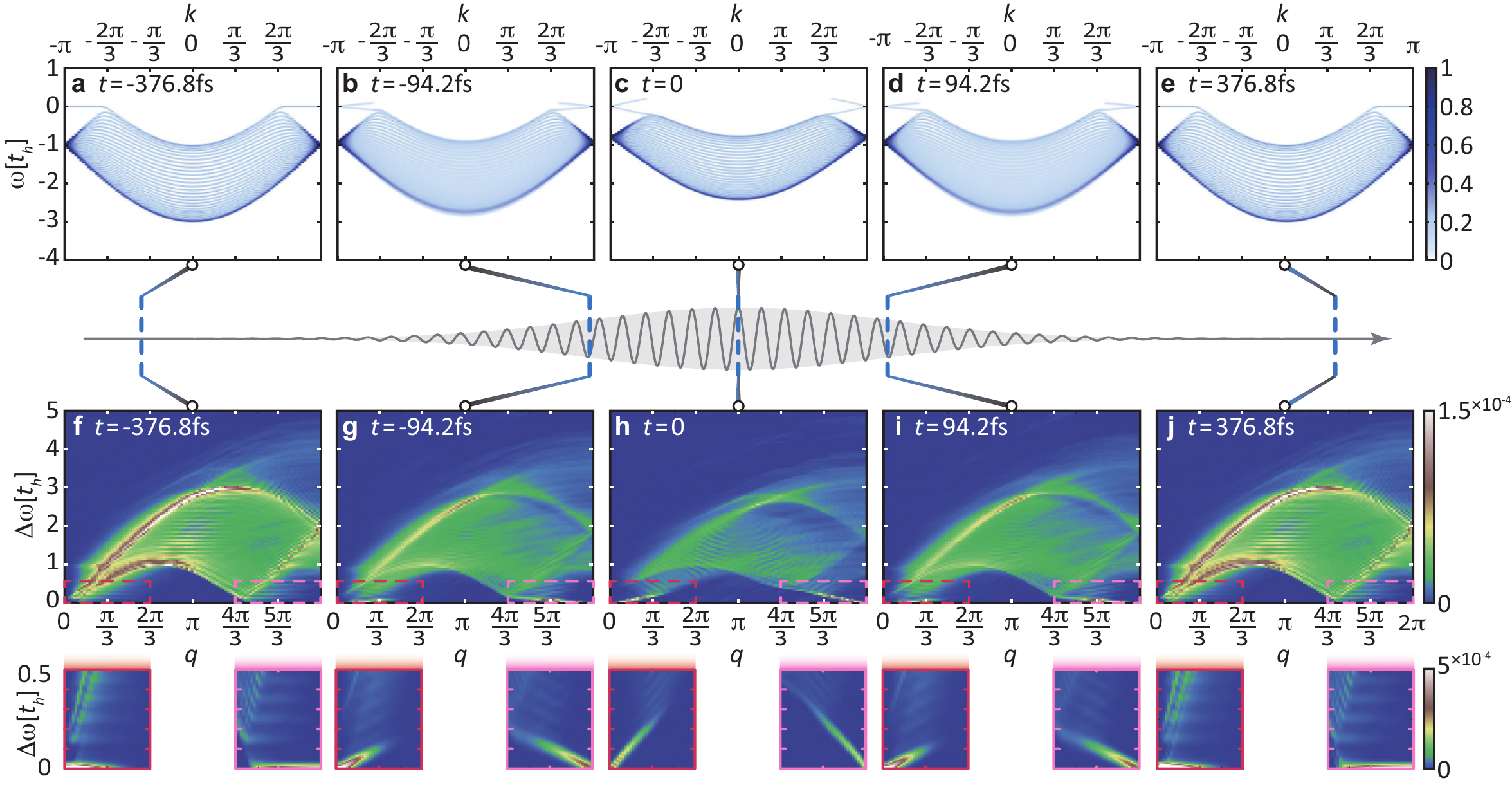}
\caption{\label{fig:2} \textbf{Snapshots of trARPES and trRIXS.} Five snapshots of trARPES (a-e) and trRIXS (f-j) are taken at (a)(f)$t=-4\sigma_{\rm pu}$, (b)(g) $t=-\sigma_{\rm pu}$, (c)(h) $t=0$, (d)(i) $t=\sigma_{\rm pu}$, and (e)(j) $t=4\sigma_{\rm pu}$ respectively. In trARPES, $k$ is the momentum along the graphene nanoribbon, $\omega$ is the energy; in trRIXS, $q$ is the momentum transfer along the graphene nanoribbon, $\Delta\omega$ is the energy loss. The middle inset sketches the temporal profile of the vector potential $\abs{\mathbf{A}(t)}$ of the circular pump, with the dashed lines representing the corresponding snapshots. The two smaller figures below each trRIXS snapshot show the zoom-in view of the regions within the red or pink boxes, highlighting the signals in the low energy region. The range of the colorbar for (a)-(e) goes from 0 to 1, while for (f)-(j) it goes from 0 to $1.5\times 10^{-4}$. For smaller figures below (f)-(j), the range of the colorbar is from 0 to $5\times 10^{-4}$.
}
\end{figure}

\begin{figure}
\includegraphics[width=1\linewidth]{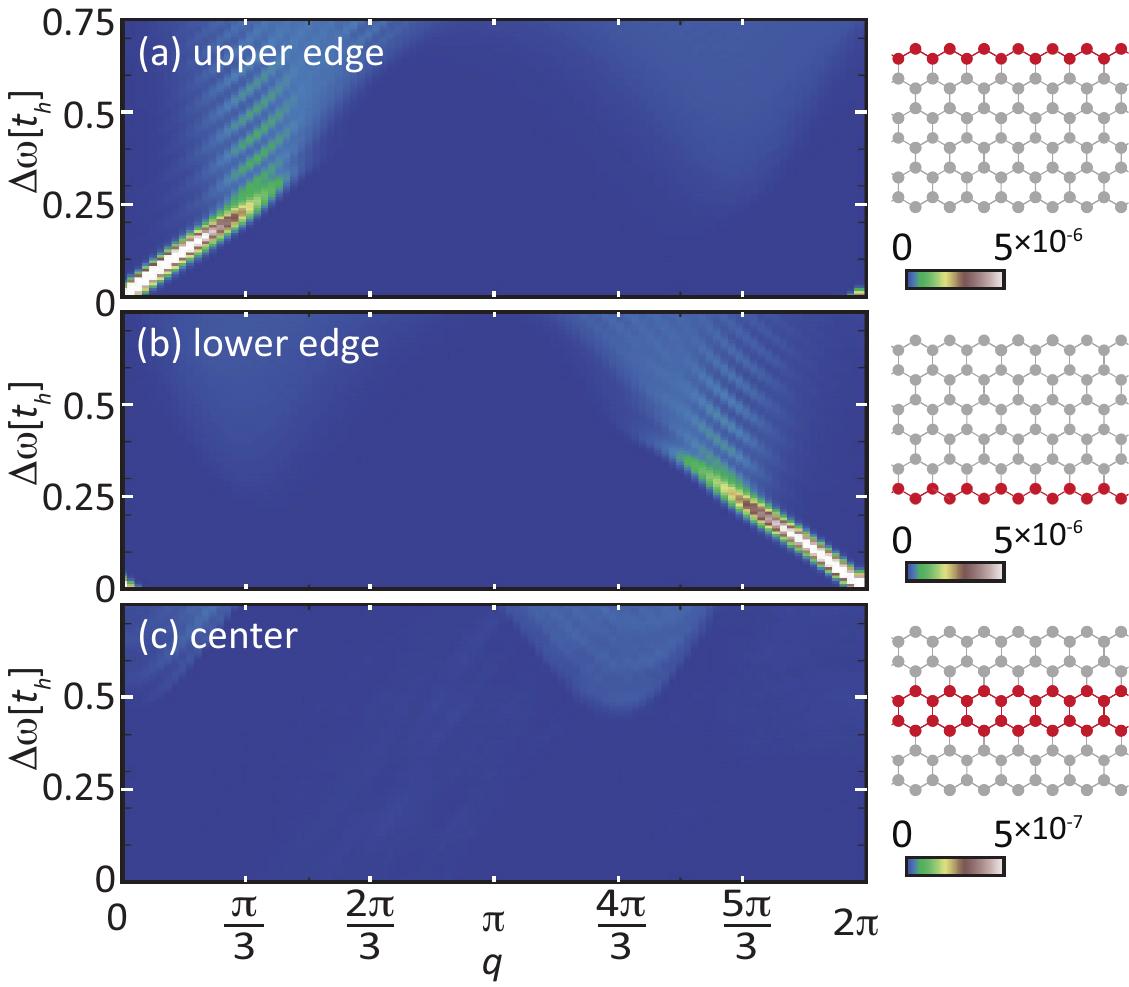}
\caption{\label{fig:3} \textbf{Spatially filtered trRIXS results.} TrRIXS signal of the graphene nanoribbon at pump center filtered on (a) upper edge ($y_0=0$), (b) lower edge ($y_0=w$) and (c) bulk ($y_0=w/2$) respectively. Here $w$ is the width of the graphene nanoribbon. The insets of (a)-(c) schematically show the filter on the graphene nanoribbon by a red part. The range of the colorbar for (a) and (b) is from 0 to $5\times 10^{-6}$; for (c) it is from 0 to $5\times 10^{-7}$.}
\end{figure}

\begin{figure}
\includegraphics[width=\linewidth]{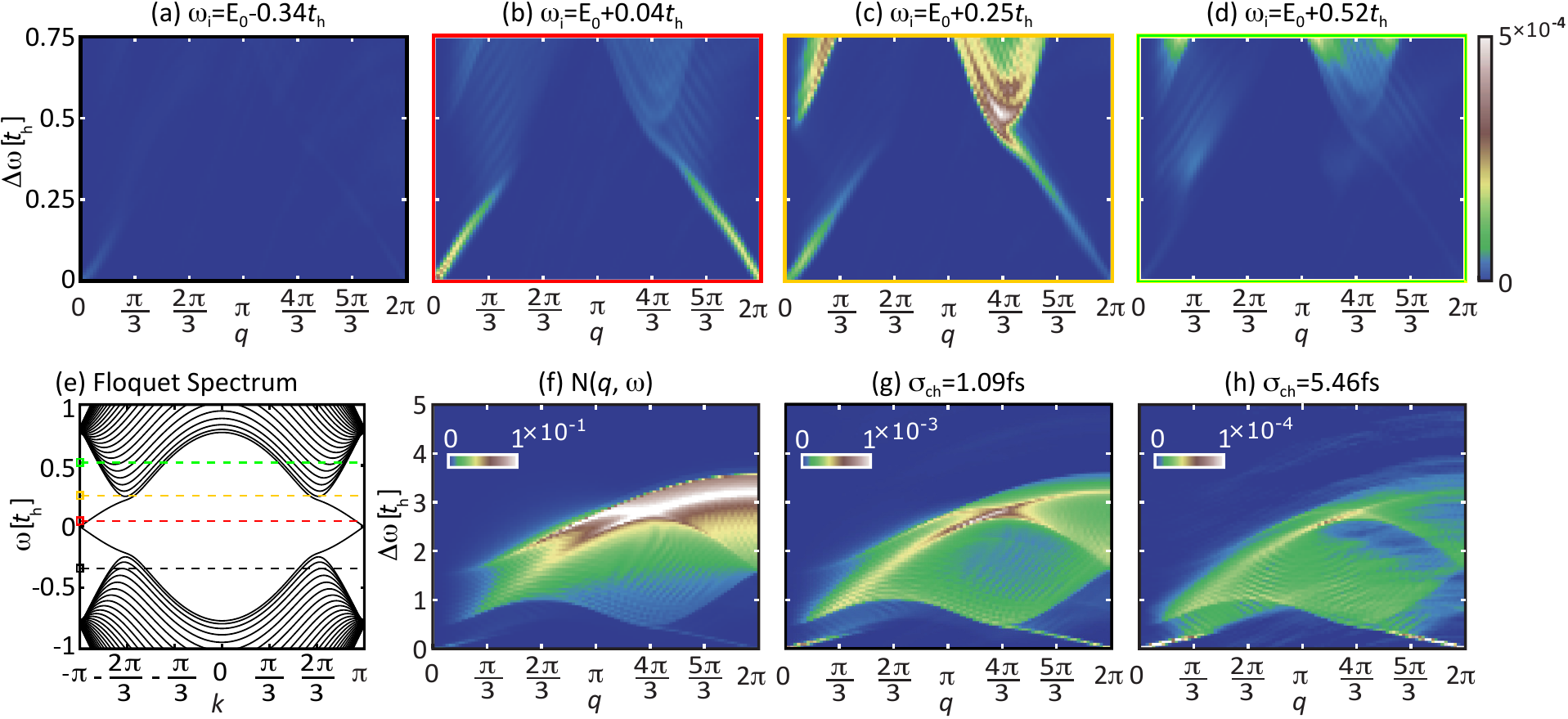}
\caption{\label{fig:4} \textbf{TrRIXS at the pump center under different incoming photon energy $\omega_{\rm i}$ and $\sigma_{\rm ch}$.} (a)-(d) The trRIXS spectrum at $t=0$ under different incoming photon energies. From (a) to (d), $\omega_{\rm i}-E_0$ is $-0.34t_h (-0.93\rm eV)$, $0.04t_h (0.12\rm eV)$, $0.25t_h(0.69\rm eV)$, and $0.52t_h (1.41\rm eV)$, respectively. They share the colorbar next to (d). Other parameters are the same as Section \ref{sec:pumpProbeRes}. (e) The corresponding Floquet spectrum of the graphene nanoribbon around the Dirac point gap. The four dashed lines from bottom to top designate different $\omega_{\rm i}-E_0$ in (a)-(d) respectively, using the colors that are the same as the corresponding frame edges. (f) The (time-dependent) charge dynamic structure factor of the graphene nanoribbon at the pump center. (g)(h) The tr-RIXS snapshots for different core-hole lifetimes $\sigma_{\rm ch}$ as shown on their titles. The range of the colorbar for (a)-(d) is 0 to $5\times 10^{-4}$. For others, it is (f) 0 to 0.1 (g) 0 to $1\times 10^{-3}$ (h) 0 to $1\times 10^{-4}$.}
\end{figure}

\begin{figure}
\includegraphics[width=\linewidth]{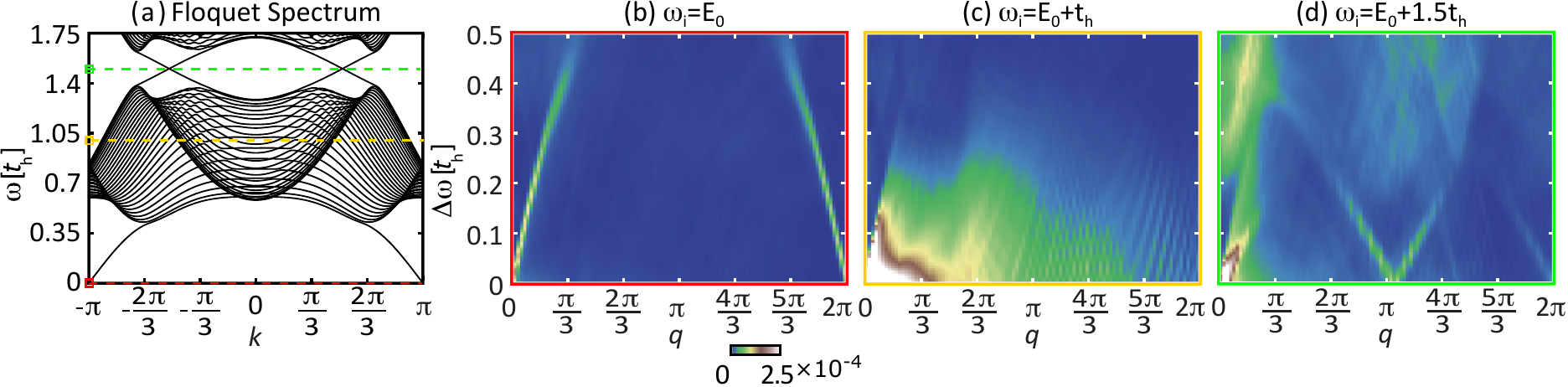}
\caption{\label{fig:5} \textbf{Low pump frequency results.} The low pump frequency results under $\Omega=3t_h$ and $A_0=0.9$. (a) The Floquet spectrum of the graphene nanoribbon under a periodic pump. The three dashed lines, from bottom to top, correspond to $\omega_{\rm i}-E_0$, which is the energy of the excited electron in the intermediate state in (b)-(d) respectively. The colors of the dashed lines match the colors of the frame edges accordingly. (b)-(d) The trRIXS spectrum at $t=0$ under different incoming photon energies. From (b) to (d), $\omega_{\rm i}-E_0$ is $0$, $t_h (2.70\rm eV)$, and $1.5t_h(4.05\rm eV)$ respectively. The range of the colorbar for (b)-(d) is 0 to $2.5\times 10^{-4}$.}
\end{figure}

We use a single-band tight-binding Hamiltonian
$\mathcal{H} = -\sum_{\langle ij \rangle} t_{ij} c_i^\dagger c_j$ on a honeycomb lattice to simulate the $2p_z$ band of the graphene nanoribbon without explicitly inclusion of various orbits, where $c_j$ annihilates an electron at site $j$. Since $2p_z$ has a different mirror symmetry by the graphene plane compared to other $2p$ and $2s$ orbitals and our pump polarization is in the graphene plane, those orbits do not mix together in such circumstances. Thus we can just take $2p_z$ orbitals when considering low energy spectra near the Fermi energy. We have also ignored the matrix element from $2p_z$ band to other bands of higher energy.
All dangling bonds are terminated by hydrogen atoms, leaving negligible contributions to the electronic states near the Fermi level $E_{\textrm{F}}$\cite{wakabayashi2010electronic}. Since the system is translational invariant along the $x$ direction, one can define a unit cell with $N$ atoms (indicated by the red box in Fig.~\ref{fig:1}):
\begin{equation}
\mathcal{C}_l = \left( c_{l,0}, c_{l,1}, \cdots, c_{l,N-1}\right)^T .
\end{equation}
Here, we relabel the coordinate of lattice by $(l,\alpha)$ as indicated in Fig.~\ref{fig:1}. We truncate the tight-binding model to only nearest-neighbor hopping $t_h=2.7\rm eV$. The calculations assume zero temperature and half filling. 

In momentum space, let $\mathcal{C}_k=\frac{1}{\sqrt{L}}\sum_l \textrm{e}^{-\textrm{i}kl}\mathcal{C}_l$, the Hamiltonian for a zigzag graphene nanoribbon can be written as
\begin{equation}\label{Hamiltonian_zigzag_original}
\thinmuskip=0mu
\medmuskip=-1mu
\begin{aligned}
\mathcal{H}(t)&=-t_h\sum_{k} \left(\sum_{\alpha=0}^{N-2}c_{k,\alpha+1}^\dagger c_{k,\alpha}+\sum_{n=0}^{\lfloor\frac{N-2}{4}\rfloor}\textrm{e}^{-\textrm{i}kD}c_{k,4n+1}^\dagger c_{k,4n}\right.\\
&\left.+\sum_{n=0}^{\lfloor\frac{N-4}{4}\rfloor}\textrm{e}^{\textrm{i}kD}c_{k,4n+3}^\dagger c_{k,4n+2}\right)+h.c.
\end{aligned}
\end{equation}
where $\lfloor\cdot\rfloor$ denotes the floor function. In this work, we adopt $N=60$ and number of unit cells as $L=100$ with periodic boundary condition along the nanoribbon. Unless otherwise specified, we take the natural unit by setting $a_0=e=\hbar=1$.


Out of equilibrium, we take the long-wavelength approximation and describe the light-matter interaction through the Peierls substitution $c_{l,\alpha}\rightarrow c_{l,\alpha}\exp[-i\mathbf{A}(t)\cdot\mathbf{r}_{l,\alpha}]$ \cite{peierls1933}. Here $\mathbf{r}_{l,\alpha}$ is the position of the carbon atom at lattice coordinate $(l,\alpha)$, and $\mathbf{A}(t)$ denotes the vector potential of the pump laser field. The Hamiltonian of a pumped graphene nanoribbon becomes
\begin{equation}\label{Hamiltonian_zigzag}
\thinmuskip=0mu
\medmuskip=-1mu
\begin{aligned}
\mathcal{H}(t)=&-t_h\sum_{k}\Bigg(\sum_{\alpha=0}^{N-2}\textrm{e}^{\textrm{i}\mathbf{A}(t)\cdot\mathbf{d}_{\alpha, \alpha+1}}c_{k,\alpha+1}^\dagger c_{k,\alpha}\\
+&\sum_{n=0}^{\lfloor\frac{N-2}{4}\rfloor}\textrm{e}^{\textrm{i}\mathbf{A}(t)\cdot\mathbf{d}_{2, 3}}\textrm{e}^{-\textrm{i}kD}c_{k,4n+1}^\dagger c_{k,4n}\\
+&\sum_{n=0}^{\lfloor\frac{N-4}{4}\rfloor}\textrm{e}^{\textrm{i}\mathbf{A}(t)\cdot\mathbf{d}_{0, 1}}\textrm{e}^{\textrm{i}kD}c_{k,4n+3}^\dagger c_{k,4n+2}\Bigg)+h.c.
\end{aligned}
\end{equation}
where $\mathbf{d}_{\alpha,\beta}=\mathbf{r}_{\beta}-\mathbf{r}_{\alpha}$ is the difference of the positions of site $\beta$ and site $\alpha$ within the same unit cell.

\subsection{Formalism for trARPES and trRIXS}
In the calculation of trARPES and trRIXS, we employ a circular-polarized pump field
\begin{equation}\label{pump}
\mathbf{A}(t)=A_0e^{-t^2/2\sigma_{\rm pu}^2}[\hat{\mathbf{e}}_x\cos(\Omega t) - \hat{\mathbf{e}}_y\sin(\Omega t)]\,,
\end{equation}
with dimensionless maximum pump amplitude $A_0=0.9$ and width of the pump pulse $\sigma_{\rm pu}=150\rm{eV}^{-1}=\textrm{98.7fs}$. For $\Omega=6.2t_h = 16.7 \rm eV$ and $\Omega=3t_h =8.1 \rm eV$, the peak pump strength $A_0=0.9$ corresponds to electric field strengths of $\mathcal{E}_0 = 10.6 \textrm{V\AA}^{-1}$ and $\mathcal{E}_0 = 5.1 \textrm{V\AA}^{-1}$, respectively, with $A_0 = e \mathcal{E}_0 a_0 / \hbar \Omega$, where $e$ is the electron charge. To resolve the transient Floquet states induced by the pump, the probe width $\sigma_{\rm pr}$ should satisfy $2\pi/\Omega \ll \sigma_{\rm pr} \ll \sigma_{\rm pu}$. Therefore, we select $\sigma_{\rm pr}=23\,\rm eV^{-1}=\textrm{15.1 fs}$ for the trARPES and trRIXS calculations.

The trARPES cross section can be written as\,\cite{FreericksPRL2009}
\begin{equation}\label{trARPES}
A(k,\omega,t)\!\propto\!-i\!\iint\! dt_1 dt_2 g(t_1;t)g(t_2;t)\textrm{e}^{\textrm{i}\omega(t_1-t_2)}G_k^< (t_2,t_1)
\end{equation}
where $G_k^< (t_2,t_1)=i\sum_{\alpha}\langle\mathcal{C}_{k,\alpha}^\dagger(t_2)\mathcal{C}_{k,\alpha}(t_1)\rangle$ is the lesser Green's function, $g(\tau;t)$ is the lineshape of the probe pulse centered at time $t$, and a proper prefactor for equation~\eqref{trARPES} is chosen to be $\sigma_{\rm pr}/\sqrt{\pi}$ (see more explanations in Supplementary Note 1). Here, we employ a Gaussian profile to mimic the realistic probe pulse
\begin{equation}\label{eq:probeShape}
g(\tau;t)=\frac1{\sqrt{2\pi}\sigma_{\rm pr}}\exp[-\frac{(\tau-t)^2}{\sigma_{\rm pr}^2}]\,.
\end{equation}

Without considering the material-specific matrix elements, the trRIXS cross-section is written as\cite{mypaper}
\begin{align}\label{trRIXS}
&I(\omega_{\rm i},\omega_{\rm f},\mathbf{q},t)\nonumber\\
=&{\int_{-\infty}^{\infty}} dt_2 {\int_{-\infty}^{t_2}} dt_1 {\int_{-\infty}^{\infty}} dt'_2 {\int_{-\infty}^{t'_2}} dt'_1 \textrm{e}^{\textrm{i}\omega_{\rm i}(t'_1-t_1)-i\omega_{\rm f}(t'_2-t_2)}\nonumber\\
&\times l(t_1,t_2)l(t'_1,t'_2)g(t_1;t)g(t_2;t)g(t'_1;t)g(t'_2;t)\nonumber\\
&\times \sum_{m,n}\textrm{e}^{\textrm{i}\mathbf{q}\cdot (\mathbf{R}_{m}-\mathbf{R}_{n})}S_{\mathbf{e}_{\rm i} \mathbf{e}_{\rm f}}^{mn}(t_1,t_2,t'_2,t'_1)
\end{align}
where $\omega_\textrm{i(f)}$ is the incoming (outgoing) photon energy, $\mathbf{q}$ is the momentum transfer, $\mathbf{R}_m$ is the lattice position at site $m$. 
The core-hole decay function $l(t_j,t_i)=\exp(-|t_j-t_i|/\sigma_{\rm ch})$ describes the lifetime effect of the core-hole induced by a resonant absorption. The four-time correlation function
\begin{equation}\label{4correlation}
\begin{aligned}
S_{\mathbf{e}_{\rm i} \mathbf{e}_{\rm f}}^{mn}(t_1,&t_2,t'_2,t'_1)=\langle U(-\infty,t'_1)\mathcal{D}_{n\mathbf{e}_{\rm i}}^\dagger U(t'_1,t'_2)\mathcal{D}_{n\mathbf{e}_{\rm f}}\\
&\times U(t'_2,t_2)\mathcal{D}_{m\mathbf{e}_{\rm f}}^\dagger U(t_2,t_1)\mathcal{D}_{m\mathbf{e}_{\rm i}} U(t_1,-\infty)\rangle
\end{aligned}
\end{equation}
depicts the multi-time correlations of resonant excitations. Here, $\mathcal{D}_{m\mathbf{e}_{\rm i}}$ is the dipole operator at site $m$ when the light polarization is labeled by $\mathbf{e}_\textrm{i}$. The detailed derivations of the trRIXS cross-section equation~\eqref{trRIXS} can be found in Ref.~\onlinecite{mypaper}. In this formalism, coherence is preserved all through. The characteristic time scales here, namely the pump period $2\pi/\Omega$ and the core-hole lifetime $\sigma_{\textrm{ch}}$, are both much smaller than the time for carriers in graphene to relax to a hot Fermi distribution, which can take up to 250 fs\cite{relax}. 

Here for the quasi-1D graphene nanoribbons, we take $\mathbf{q}=(q,0)$ and scan $q$ along the direction of the ribbon. Under equation~\eqref{Hamiltonian_zigzag},  equation~\eqref{trRIXS} can be simplified using
\begin{equation}\label{ribbon_RIXS}
\begin{aligned}
&\sum_{m,n}\textrm{e}^{\textrm{i}\mathbf{q}\cdot (\mathbf{R}_{m}-\mathbf{R}_{n})}S_{\mathbf{e}_{\rm i} \mathbf{e}_{\rm f}}^{mn}(t_1,t_2,t'_2,t'_1)\\
=&\sum_{\alpha,\beta}\textrm{e}^{\textrm{i}\mathbf{q}\cdot (\mathbf{r}_{\alpha}-\mathbf{r}_{\beta})}\langle \mathcal{C}_{k+q,\beta}(t_1') \mathcal{C}_{k,\beta}^\dagger (t_2') \mathcal{C}_{k,\alpha}(t_2) \mathcal{C}_{k+q,\alpha}^\dagger(t_1) \rangle
\end{aligned}
\end{equation}
Here we have implicitly assumed that the kinetic energy of the core-hole is negligible at the ultrafast timescale and therefore only kept two site indices $(\alpha,\beta)$ in each term. The mathematical evaluation of equation~\eqref{ribbon_RIXS} is explained in Supplementary Note 3.
\subsection{Spatial Filters of trRIXS Features}
A spatial filter is added to the graphene nanoribbon by substituting equation~\eqref{ribbon_RIXS} with
\begin{equation}\label{nano_RIXS}
\sum_{m,n}{\rm e}^{{\rm i}\bf{q}\cdot (\bf{r}_{\alpha}-\bf{r}_{\beta})}e^{-\frac{(\mathbf{r}_{\alpha}\cdot \mathbf{e}_y - y_0)^2}{2\sigma_{r}^2}-\frac{(\mathbf{r}_{\beta}\cdot \mathbf{e}_y - y_0)^2}{2\sigma_{r}^2}}S_{\mathbf{e}_{\rm i} \mathbf{e}_{\rm f}}^{mn}(t_1,t_2,t'_2,t'_1)
\end{equation}
where $y_0$ is the position of the filter center along $y$ axis, $\mathbf{e}_y$ is the unit vector along the $y$ axis, $\sigma_r$ is the width of the filter which is taken to be 2.
\subsection{Reducing TrRIXS to Charge Dynamic Structure Factor}
In an extreme case of UCL, the core-hole lifetime $\sigma_{\rm ch}=0$. Then in equation~\eqref{trRIXS}, we take $l(t_i,t_j)=\delta(t_i-t_j)$. Equation~\eqref{trRIXS} becomes
\begin{equation}\label{simplified_RIXS}
\begin{aligned}
&I(\Delta\omega,\mathbf{q},t)\\
=&{\int_{-\infty}^{\infty}} dt_1 {\int_{-\infty}^{\infty}} dt'_1 \textrm{e}^{\textrm{i}\Delta\omega(t'_1-t_1)}s(t_1,t)s(t'_1,t)\\
&\times \sum_{k,\alpha,\beta}\textrm{e}^{\textrm{i}\mathbf{q}\cdot (\mathbf{r}_{\alpha}-\mathbf{r}_{\beta})}\langle \mathcal{C}_{k+q,\beta}(t_1') \mathcal{C}_{k,\beta}^\dagger (t_1') \mathcal{C}_{k,\alpha}(t_1) \mathcal{C}_{k+q,\alpha}^\dagger(t_1) \rangle
\end{aligned}
\end{equation}
where $s(t_1,t) = g^2(t_1,t)$. Define the charge density $\rho_{\mathbf{q}}(t_i)=\sum_{k,\alpha}\exp(\textrm{i}\mathbf{q}\cdot\mathbf{r}_\alpha)\mathcal{C}^\dagger_{k+q,\alpha}(t_i)\mathcal{C}_{k,\alpha}(t_i)$. Then the trRIXS cross section is just the time-dependent charge dynamic structure factor
\begin{equation}
\begin{aligned}
N(\mathbf{q},\omega,t)={\int_{-\infty}^{\infty}} dt_1 {\int_{-\infty}^{\infty}} dt'_1 &\textrm{e}^{\textrm{i}\omega(t'_1-t_1)}s(t_1,t)s(t'_1,t)\\
&\times\langle \rho_{\mathbf{q}}(t'_1) \rho_{-\mathbf{q}}(t_1) \rangle
\end{aligned}
\end{equation}.

\section*{Data Availability}
The data that support the findings of this study are available from the corresponding author upon reasonable request.

\section*{Code Availability}
The computer codes used in this study will be made available upon reasonable requests to the corresponding author.

\section*{Acknowledgements}
The authors thank T. Cao, M. Sch\"{u}ler, and Y. He for helpful discussions. The work at Stanford University and SLAC National Accelerator Laboratory was supported by the US Department of Energy, Office of Basic Energy Sciences, Division of Materials Sciences and Engineering, under Contract No. DE-AC02-76SF00515.  M. C. acknowledges support from the Flatiron Institute Center for Computational Quantum Physics. The Flatiron Institute is a divison of the Simons Foundation.  This research used resources of the National Energy Research Scientific Computing Center (NERSC), a US Department of Energy Office of Science User Facility operated under Contract No. DE-AC02-05CH11231.

\section*{Competing Interests}
The authors declare no competing interests.

\section*{Author Contributions}
Y.C., Y.W., and T.P.D designed the research. Y.C., Y.W., and M.C. conducted the theoretical derivations, while Y.C. conducted the numerical calculations and data analysis. All authors discussed results and contributed to writing the manuscript.




\clearpage

\clearpage
\end{document}